# Upconversion mid-infrared dual-comb spectroscopy

## Zaijun CHEN,[1] Theodor W. HÄNSCH,[1,2] and Nathalie PICQUÉ[1,*]

[1] *Max-Planck Institute of Quantum Optics, Hans-Kopfermann-Str. 1, 85748 Garching, Germany*
[2] *Ludwig-Maximilian University of Munich, Faculty of Physics, Schellingstr. 4/III, 80799 Munich, Germany*
*Corresponding author: nathalie.picque@mpq.mpg.de*

**Abstract: We devise a new detection technique for mid-infrared multi-heterodyne spectroscopy. As an experimental implementation, mid-infrared light interrogates a gas sample in the 3-µm region of the fundamental CH, OH, NH stretch in molecules and detection is performed in the near-infrared telecommunication region. Spectra showing 18000 resolved comb lines of 100-MHz spacing are recorded at a sensitivity close to the shot-noise limit. The demonstration adds to the unique advantages of dual-comb spectroscopy, such as potential very high resolution, negligible instrumental line width, and direct calibration of the frequency scale with an atomic clock. We also show that direct mid-infrared detection with a differential photo-detector module is beneficial in the limit of low mid-infrared average power.**

We explore a scheme of mid-infrared dual-comb spectroscopy with detection in the near-infrared region. We therefore take advantage of the strong line intensity of the fundamental ro-vibrational transitions of molecules in the mid-infrared and of the sensitive detection in the near-infrared.

The mid-infrared spectral range (2-20 µm) is important for molecular spectroscopy and sensing: the molecular fingerprint region is the domain of strong vibrational transitions in molecules. In gas-phase spectroscopy, the intensity of the fundamental ro-vibrational lines can be more than a hundred-fold stronger than their overtones in the near-infrared region. The sensitivity to weak absorption may thus be significantly improved. Unfortunately, mid-infrared detectors are usually not as sensitive and fast as near-infrared ones.

Therefore, numerous implementations in imaging, in spectroscopy and in frequency metrology rely on upconversion [1-5]: the mid-infrared light interrogates the sample and is converted to the near-infrared or visible spectral range for detection.

The recent technique of dual-comb spectroscopy currently attracts significant interest [6], for it enables a distinctive combination of features: it interrogates broad spectral bandwidths with a single photo-detector, providing a high consistency -characteristic of multiplex spectroscopic techniques- and the potential ability to be implemented in any spectral region, from sub-millimeter waves to the extreme ultraviolet. The frequency scale of the spectra can be calibrated to an atomic clock and the signature of the instrument -the instrumental line-shape- can be neglected for transitions of small molecules at Doppler-broadened room-temperature conditions. Concurrently to the advance of mid-infrared frequency comb technology [7], the development of dual-comb spectroscopy in the mid-infrared spectral range has been reported in several publications with fully-stabilized frequency comb generators [8-10] or with simpler systems [11-17], that do not leverage all the potential advantages of dual-comb spectroscopy. Reaching the highest possible sensitivities would increase the applicability of mid-infrared dual-comb spectroscopy. Upconversion techniques have already been explored with electro-optic sampling in the THz [18, 19] and deep mid-infrared [20] regions, although the possible sensitivity increase has not been quantified yet. Here, we propose a new scheme of upconversion dual-comb spectroscopy. We experimentally demonstrate the concept in a configuration based on difference frequency generation of a femtosecond oscillator and a continuous-wave laser, although our general scheme could be advantageously applied with frequency comb generators that do not emit ultra-short pulses, such as quantum cascade lasers.

Our dual-comb set-up for upconversion mid-infrared spectroscopy is sketched in Fig. 1a. A self-referenced amplified erbium-doped fiber frequency comb synthesizer (called master comb generator) emits pulses at a repetition frequency of $f_{rep}$=100 MHz. The repetition frequency and the carrier-envelope offset frequency of the master comb are stabilized against the radio-frequency signal of an active hydrogen maser, which also serves to synchronize all electronics in the experiment. The width of the optical lines of the master comb are less than 100 kHz at 1-minute averaging. One output beam of the master comb oscillator, with a span of more than 20 THz and a center frequency of 190 THz, is used for spectroscopy. It is converted to the mid-infrared 92-THz (3.3-μm) region by difference frequency generation in a 20-mm-long periodically poled lithium niobate (PPLN) crystal. The pump beam for difference frequency generation is delivered by a 281.8-THz continuous-wave laser with a narrow line-width (<50 kHz). The pump laser is phase-locked to one line of the comb. An average power of 250 mW of the master comb oscillator is combined to 4 W of the pump laser. An idler spectrum comprising 18000 comb lines and spanning about 1.8 THz is generated. The center frequency can be tuned between 88 and 102 THz. The span is limited by the phase matching conditions in the PPLN crystal. Tuning of the center frequency is achieved by changes of temperature and poling-period. Due to the uneven spectral radiance of the near-infrared comb, the mid-infrared beam has an average power varying between 230 μW at a center frequency of 90.5 THz to 700 μW at a center frequency of 92.6 THz. In the frequency range 600 kHz - 10 MHz, the experimentally measured relative intensity noise of the mid-infrared comb is constant and equal to -144 dBc·Hz$^{-1}$. The mid-infrared beam passes through a cell of a length of 2 cm and of a pressure of 400 Pa of methane in natural abundance. The mid-infrared comb is then converted back to the telecommunication region using difference frequency generation with a beam of the same continuous-wave laser, at an average power of 2.4 W, and a second 20-mm-long PPLN crystal. Using the same continuous-wave laser for down- and upconversion re-maps the comb lines at their initial frequency and facilitates the heterodyning with a second comb. The continuous-wave laser also greatly simplifies the implementation, as the temporal overlap with the mid-infrared pulses does not require any active or passive stabilization scheme. The near-infrared upconverted comb has an average power of 400 nW when the mid-infrared center frequency is 90.5 THz, and of 1.3 µW when converted from 92.6 THz. At 92.6 THz, the relative intensity noise of the near-infrared upconverted comb is -142.5 dBc·Hz$^{-1}$ in the frequency range 600 kHz - 10 MHz.

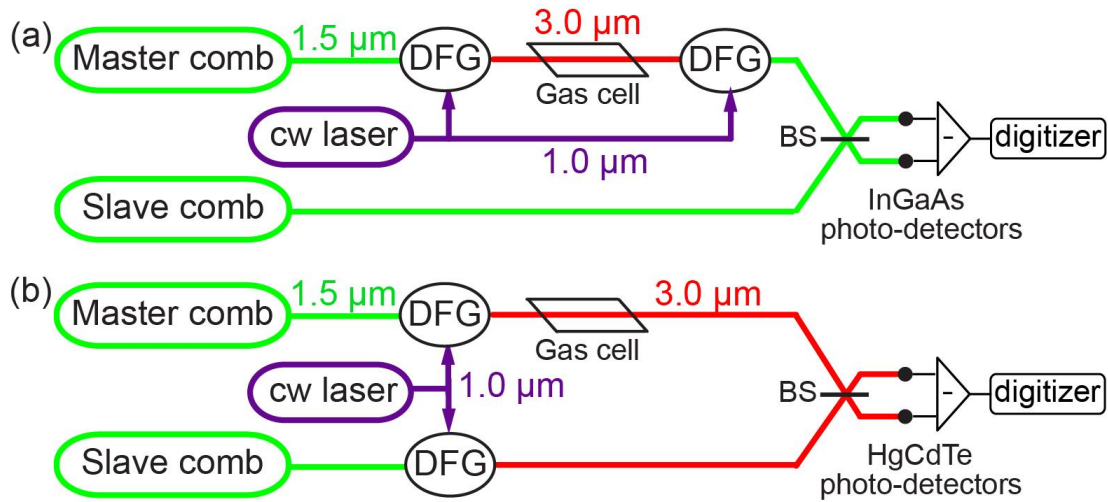

**Figure 1.** (a) Experimental set-up for upconversion mid-infrared dual-comb spectroscopy (b) experimental set-up for mid-infrared dual-comb spectroscopy with direct mid-infrared differential detection. DFG: difference frequency generation. cw: continuous-wave. BS: beam-splitter.

The upconverted beam is combined with that of a second erbium comb, called slave comb generator, of slightly different repetition frequency $f_{rep}+\Delta f_{rep}$ with $\Delta f_{rep}$=200 Hz. The coherence between the master comb and the slave comb is enforced, before difference frequency generation, by feed-forward control of the relative carrier-envelope offset frequency, as described in [8, 21]. The relative carrier-envelope frequency $\Delta f_{ceo}$ is set to zero (modulo $\Delta f_{rep}$), as this is one of the conditions to produce reproducible interferometric waveforms that enable direct time-domain averaging. In the 3-µm region, the scheme of feed-forward dual-comb spectroscopy has demonstrated efficient direct averaging of the interferograms, without requiring any analog or digital data corrections, over times longer than 30 minutes. The slave comb is spectrally filtered using a pair of gratings so that its bandwidth matches that of the upconverted comb. The optically-filtered slave comb has a relative intensity noise of -149 dBc·Hz$^{-1}$ in the frequency range 600 kHz - 10 MHz. Its beam is attenuated to avoid detector nonlinearities: its average power does not exceed 20 µW onto each InGaAs detector of a balanced differential detection module which subtracts the two outputs of the interferometer. Around a radio-frequency of 4 MHz, the noise equivalent power of the InGaAs balanced-detection module is measured to 0.9 pW·Hz$^{-1/2}$. The time-domain interference signal is digitized with a data acquisition board with a sample rate of 100x10$^6$ samples s$^{-1}$. The effective number of bits of the data acquisition board is 12 bits. The radio-frequency comb is mapped into the range of 1-5 MHz.

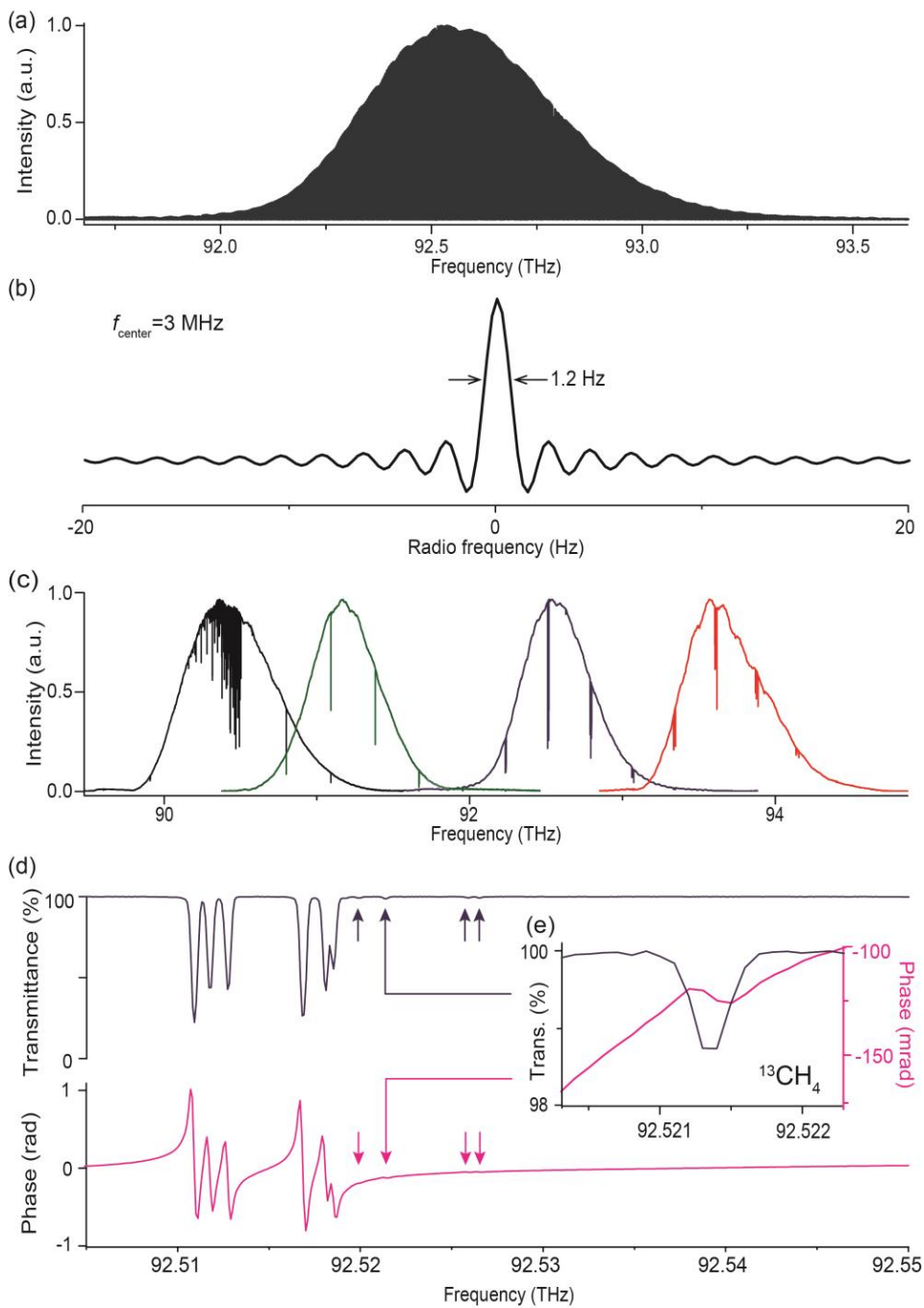

**Figure 2.** Experimental upconversion dual-comb spectra. (a) Apodized spectrum over the entire span of 1.8 THz. The spectrum results for 1000 averaged interferograms, each of 1-s measurement time. (b) Comb line in the unapodized spectrum, centered at 3 MHz in the radio-frequency spectrum. A perfect cardinal-sine shape is observed with a full-width at half maximum of 1.2 Hz in the radio-frequency domain, corresponding to 1-second measurement time [22]. (c) Four experimental dual-comb spectra sampled at exactly the comb line spacing of 100 MHz in the region of the $Q$-branch and the $R$-branch of the $\nu_3$ band of $^{12}CH_4$. (d) Portion of experimental upconversion transmittance (blue) and phase (magenta) spectra showing the strong $R(6)$ manifold of the $\nu_3$ band of $^{12}CH_4$ at 100-MHz resolution. The weak lines between 92.52 and 93.53 THz (indicated with arrows) belong to the $R(7)$ manifold of the $\nu_3$ band of $^{13}CH_4$, [23]. (e) Magnified representation of (d) with the transmittance and phase spectra of the $A_2(0)$ tetrahedral sub-level line in the $R(7)$ manifold at 92.5214 THz.

An experimental apodized spectrum (Fig.2a) spans 1.8 THz with 18000 resolved comb lines at a center frequency of 92.6 THz. Its measurement time is 1000 seconds: one thousand interferograms, each of a duration of 1 second, are averaged. The averaged interferogram is four-fold zero filled to interpolate the spectrum [22] and it is Fourier transformed. The amplitude of the complex Fourier transform is shown with triangular apodization in Fig 2a, with the frequency scale converted to its mid-infrared values. An unapodized representation of one comb line is displayed in Fig. 2b on a radio-frequency scale. The comb line has a cardinal sine shape, which is the expected instrument signature. As the measurement time is finite, the interferogram is multiplied by a boxcar function, and in the spectrum, each comb line is convolved by a cardinal sine. The full-width at half-maximum of the comb lines is 1.2 Hz in the radio-frequency domain, which corresponds exactly to the Fourier transform limit [22]. The shape of the comb lines illustrates the efficiency of our time-domain averaging technique, which also confirmed by the fact that the signal-to-noise ratio in the spectrum evolves with the square-root of the averaging time. Resolving the comb lines is useful for assessing subtle instrumental artifacts. Because, in addition, the optical lines of the interrogating comb are narrow and with accurate positions, the spectrometer has a high frequency accuracy and an instrumental line-shape that can be neglected for the measurement of Doppler profiles.

Once the instrument is characterized though, computing radio-frequency spectra that are exactly sampled at the comb line positions, of spacing $\Delta f_{rep}$, is equivalent and sufficient. Four such experimental dual-comb spectra, at an optical resolution equal to the comb line spacing $f_{rep}$=100 MHz, are shown (Fig.2c) in the region of the $Q$-branch and the $R$-branch of the $\nu_3$ band of methane $^{12}$CH$_4$. The spectra, centered at 90.3 THz, 91.2 THz, 92.6 THz (Fig.2d,2e), and 93.6 THz, respectively, are recorded within 1000 s each. The signal-to-noise ratio (defined as the ratio of the signal to the standard-deviation of the fluctuations of the baseline) in the spectrum centered at 90.3 THz (also in Fig.3a) culminates at 8200. The average signal-to-noise ratio across a 1.5-THz span (defined as the spectral range for which the signal-to-noise ratio is higher than 100), from 89.8 THz to 91.3 THz, is 2540. The figure-of-merit, defined as the product of the average signal-to-noise in unit measurement time times the number of spectral elements, is 1.2×10$^6$ Hz$^{1/2}$. The experimental figure-of-merit is 1.4 times above the shot-noise-limited figure-of-merit. The relative intensity noise of the upconverted comb and the detector noise are the main other contributors, whereas the relative intensity noise of the slave comb and the digitization noise are negligible. For the calculation of the different contribution to the noise, we use the formalism initially developed in [24] and adapted in [25] to dual-comb spectroscopy.

The signal-to-noise ratio in the transmittance spectrum centered at 92.6 THz culminates at 12400 around 92.54 THz (Fig.2d). The average signal-to-noise ratio across a 1.8-THz span (91.7-93.5 THz) is 4000, which leads to a figure of merit of 2.3×10$^6$ Hz$^{1/2}$. Here the experimental figure-of-merit is 1.8 times above the shot-noise limit and the main contributions to the noise figure are the shot noise followed by the relative intensity noise of the upconverted comb. Fig. 2d also displays a portion of the dispersion (phase) spectrum.

Owing to the high signal-to-noise ratio in the transmittance and phase spectra, the transitions in the $^{13}$CH$_4$ isotopologue [23], of a natural abundance of 1.1%, are clearly measurable, even when they are in the wings of strong dispersive lines in the phase spectrum (Fig.2e).

Our experimental figures-of-merit are higher than those of reports [8, 9, 20] of fiber-laser-based mid-infrared high-resolution dual-comb spectroscopy in conditions where the detector nonlinearities are controlled. For the highest power of the upconverted comb (1.3 μW when converted from 92.6 THz), it is similar to what was obtained [12] over a narrow spectral bandwidth (1800 comb lines spanning 207 GHz) with electro-optic modulators and detection in the mid-infrared.

For a quantitative comparison, experiments with direct mid-infrared detection are performed (Fig.1b). The master comb and the slave comb are both converted to the 3-μm region using 20-mm-long PPLN crystals, exactly as the master comb was in the upconversion set-up of Fig. 1a. The settings of difference in repetition frequencies $\Delta f_{rep}$ and in carrier-envelope offset frequencies $\Delta f_{ceo}$ are kept identical as those with the upconversion detection. The optical spectra are mapped in the same 1-5 MHz radio-frequency range. The span and the power of the idler beams are the same as those of the master idler beam in the upconversion experiment reported above. The master comb idler interrogates the gas cell. The idler beams from the master comb and from the slave comb are combined on a 50/50 beam splitter.

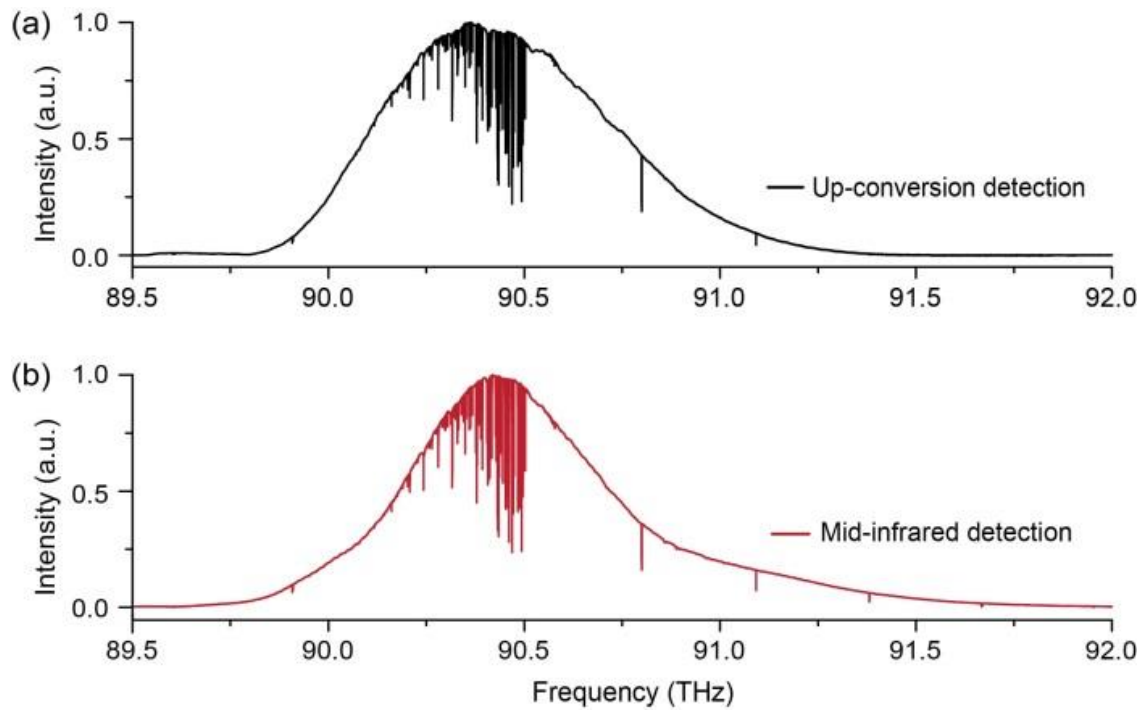

**Figure 3**. Experimental dual-comb amplitude spectrum of the *Q*-branch of the $\nu_3$ band of $^{12}$CH$_4$ with (a) upconversion near-infrared detection and (b) mid-infrared differential detection. The interferograms are directly averaged in the time domain during 1000 seconds, without any corrections, and the amplitude of the Fourier transform is represented in the figure, sampled at exactly the comb line spacing of 100 MHz (which also sets the spectral resolution).

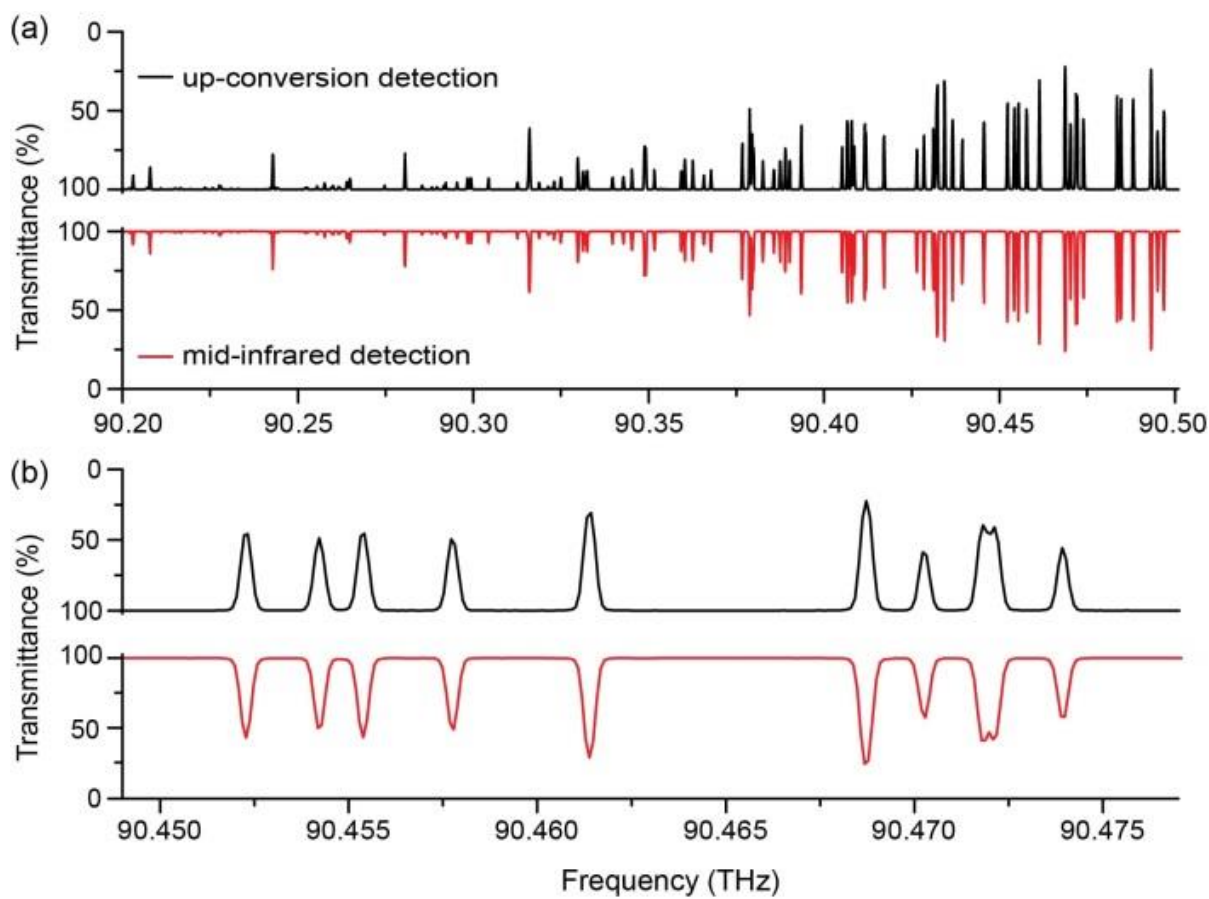

**Figure 4.** Experimental transmittance spectra recorded with the two detection schemes. (a) A portion of the transmittance spectrum in the region of the *Q*-branch of the $\nu_3$ band of $^{12}CH_4$. (b) Magnified representation of (a) around 90.46 THz with the molecular lines from the *Q*(4)-*Q*(6) manifolds of the $\nu_3$ band of $^{12}CH_4$.

When we detect one output of the interferometer with the same single HgCdTe detector as in our previous publication [8], we obtain the same figure-of-merit of $1.1 \times 10^6$ $Hz^{1/2}$ as in [8] when the spectral span was broader (80000 comb lines spaced by 100MHz instead of 18000 modes here), showing that the metric is a relevant quality factor.

Next we implement a differential detection of the two output ports of the interferometer (Fig.1b). Two identical thermoelectrically-cooled HgCdTe detectors collect the optical signals. To avoid nonlinearities, the total power onto each HgCdTe detector is attenuated to 30 μW (15 μW from each comb). Each detector is amplified with a trans-impedance amplifier and the two amplified outputs are subtracted using a differential amplifier. The 3-dB frequency bandwidth of the entire HgCdTe differential-detection module is 40 MHz and its common-mode rejection ratio in the frequency range 0-10 MHz is 43 dB. The noise equivalent power of the detection module is 9 pW·$Hz^{-1/2}$, an order of magnitude larger than that of the InGaAs balanced detector.

Figure 3 provides a comparison of two spectra, one (Fig.3a) measured with the upconversion scheme of Fig. 1a and already described above, one (Fig.3b) with direct mid-infrared differential detection (sketched in Fig.1b) The signal-to-noise ratio of the spectrum recorded with the mid-infrared balanced detection in Fig. 3b exceeds 13300 at around 90.3 THz, with an average signal-to-noise ratio of 4120 over the span of 2.2 THz (from 89.7 THz to 91.9 THz) at a total measurement time of 1000 seconds. The corresponding experimental figure-of-merit is $2.9 \times 10^6 \, Hz^{1/2}$. The strongly dominant contribution to the figure-of-merit is the detector noise. These results represent, to our knowledge, the highest signal-to-noise ratio and figure-of-merit for self-referenced mid-infrared dual-comb spectroscopy in conditions where the detector nonlinearities are circumvented. The intensity of the spectral envelope with direct mid-infrared detection (Fig. 3b) decreases more quickly than that with upconverted near-infrared detection (Fig.3a) because the two mid-infrared combs have the same bell-shaped envelope whereas the near-infrared local oscillator is rather flat-top. Even in this unfavorable situation, the figure-of-merit is higher with mid-infrared differential detection than with upconverted near-infrared detection. The figure-of-merit is constant over the tunability range of the idler combs. Figure 4 shows expanded portions of the transmittance spectra of Fig.3 with flattened baseline in the region of the $Q$-branch of the $\nu_3$ band of $^{12}CH_4$. It illustrates the high signal-to-noise ratio in both configurations as well as the good agreement between the two spectra.

Our upconversion mid-infrared spectrometer demonstrates an attractive signal-to-noise ratio, close to shot-noise-limited detection. The scheme has the advantage not to require ultrashort pulses and therefore it could be used with various types of frequency comb generators including quantum cascade lasers or electro-optic modulators. With ultra-short pulse lasers, high conversion powers could be obtained by strongly chirping the pulses so to maximize the temporal overlap with the continuous-wave laser Future work includes an assessment deeper in the mid-infrared region (8-12µm), where the detectors are less sensitive. Furthermore, in particular circumstances of a very low mid-infrared average power, a mid-infrared differential detector brings an unparalleled sensitivity and provides a straightforward mean of improving the signal-to-noise ratio in mid-infrared dual-comb spectroscopy. All these developments are shown to be entirely compatible with the important features of feed-forward dual-comb spectroscopy [8, 21]: direct time-domain averaging of the interferograms without digital corrections, high resolution, high accuracy, narrow instrumental line-shape. They will be part of new strategies for precision mid-infrared spectroscopy.

**Funding.** Carl-Friedrich-von-Siemens Foundation. Max-Planck-Fraunhofer cooperations. Max-Planck Society.

# REFERENCES

1.     A. Barh, P. J. Rodrigo, L. Meng, C. Pedersen, and P. Tidemand-Lichtenberg, "Parametric upconversion imaging and its applications," Adv. Opt. Photon. **11**, 952-1019 (2019).

2.     J. S. Dam, P. Tidemand-Lichtenberg, and C. Pedersen, "Room-temperature mid-infrared single-photon spectral imaging," Nature Photonics **6**, 788-793 (2012).

3.      S. Junaid, S. Chaitanya Kumar, M. Mathez, M. Hermes, N. Stone, N. Shepherd, M. Ebrahim-Zadeh, P. Tidemand-Lichtenberg, and C. Pedersen, "Video-rate, mid-infrared hyperspectral upconversion imaging," Optica **6**, 702-708 (2019).

4.      T. W. Neely, L. Nugent-Glandorf, F. Adler, and S. A. Diddams, "Broadband mid-infrared frequency upconversion and spectroscopy with an aperiodically poled LiNbO3 waveguide," Opt. Lett. **37**, 4332-4334 (2012).

5.      R. Santagata, D. B. A. Tran, B. Argence, O. Lopez, S. K. Tokunaga, F. Wiotte, H. Mouhamad, A. Goncharov, M. Abgrall, Y. Le Coq, H. Alvarez-Martinez, R. Le Targat, W. K. Lee, D. Xu, P. E. Pottie, B. Darquié, and A. Amy-Klein, "High-precision methanol spectroscopy with a widely tunable SI-traceable frequency-comb-based mid-infrared QCL," Optica **6**, 411-423 (2019).

6.      N. Picqué and T. W. Hänsch, "Frequency comb spectroscopy," Nature Photonics **13**, 146-157 (2019).

7.      A. Schliesser, N. Picqué, and T. W. Hänsch, "Mid-infrared frequency combs," Nature Photonics **6**, 440-449 (2012).

8.      Z. Chen, T. W. Hänsch, and N. Picqué, "Mid-infrared feed-forward dual-comb spectroscopy," Proceedings of the National Academy of Sciences **116**, 3454-3459 (2019).

9.      G. Ycas, F. R. Giorgetta, E. Baumann, I. Coddington, D. Herman, S. A. Diddams, and N. R. Newbury, "High-coherence mid-infrared dual-comb spectroscopy spanning 2.6 to 5.2 mu m," Nat Photonics **12**, 202-208 (2018).

10.     A. V. Muraviev, V. O. Smolski, Z. E. Loparo, and K. L. Vodopyanov, "Massively parallel sensing of trace molecules and their isotopologues with broadband subharmonic mid-infrared frequency combs," Nature Photonics **12**, 209-214 (2018).

11.     G. Villares, A. Hugi, S. Blaser, and J. Faist, "Dual-comb spectroscopy based on quantum-cascade-laser frequency combs," Nature Communications **5**, 5192 (2014).

12.     M. Yan, P.-L. Luo, K. Iwakuni, G. Millot, T. W. Hänsch, and N. Picqué, "Mid-infrared dual-comb spectroscopy with electro-optic modulators," Light: Science & Applications **6**, e17076 (2017).

13.     M. Yu, Y. Okawachi, A. G. Griffith, N. Picqué, M. Lipson, and A. L. Gaeta, "Silicon-chip-based mid-infrared dual-comb spectroscopy," Nature Communications **9**, 1869 (2018).

14.     L. A. Sterczewski, J. Westberg, C. L. Patrick, C. Soo Kim, M. Kim, C. L. Canedy, W. W. Bewley, C. D. Merritt, I. Vurgaftman, J. R. Meyer, and G. Wysocki, "Multiheterodyne spectroscopy using interband cascade lasers," Optical Engineering **57**, 011014 (2018).

15.     F. Zhu, A. Bicer, R. Askar, J. Bounds, i. A. A. Kolomenski, V. Kelessides, M. Amani, and H. A. Schuessler, "Mid-infrared dual frequency comb spectroscopy based on fiber lasers for the detection of methane in ambient air," Laser Physics Letters **12**, 095701 (2015).

16.     Y. Jin, S. M. Cristescu, F. J. M. Harren, and J. Mandon, "Femtosecond optical parametric oscillators toward real-time dual-comb spectroscopy," Applied Physics B **119**, 65-74 (2015).

17.     O. Kara, L. Maidment, T. Gardiner, P. G. Schunemann, and D. T. Reid, "Dual-comb spectroscopy in the spectral fingerprint region using OPGaP optical parametric oscillators," Optics Express **25**, 32713-32721 (2017).

18.     G. Klatt, R. Gebs, C. Janke, T. Dekorsy, and A. Bartels, "Rapid-scanning terahertz precision spectrometer with more than 6 THz spectral coverage," Optics Express **17**, 22847-22854 (2009).

19.     I. A. Finneran, J. T. Good, D. B. Holland, P. B. Carroll, M. A. Allodi, and G. A. Blake, "Decade-Spanning High-Precision Terahertz Frequency Comb," Physical Review Letters **114**, 163902 (2015).

20.     A. S. Kowligy, H. Timmers, A. J. Lind, U. Elu, F. C. Cruz, P. G. Schunemann, J. Biegert, and S. A. Diddams, "Infrared electric field sampled frequency comb spectroscopy," Science Advances **5**, eaaw8794 (2019).

21.     Z. Chen, M. Yan, T. W. Hänsch, and N. Picqué, "A phase-stable dual-comb interferometer," Nature Communications **9**, 3035 (2018).

22.     P. R. Griffiths, J. A. De Haseth, Fourier Transform Infrared Spectroscopy. (John Wiley & Sons Inc., Hoboken, New Jersey, ed. 2nd, 2007), pp. 529.

23.     M. Dang-Nhu, G. Poussigue, G. Tarrago, A. Valentin, and P. Cardinet, "Étude de la bande v3 de 13CH4 entre 2 863 et 3132 cm-1," Journal de Physique France **34**, 389-401 (1973).

24.     S. P. Davis, M. C. Abrams, and J. W. Brault, *Fourier Transform Spectrometry* (Academic Press, San Diego, 2001), pp. 1- 262.

25.     N. R. Newbury, I. Coddington, and W. Swann, "Sensitivity of coherent dual-comb spectroscopy," Optics Express **18**, 7929-7945 (2010).